\documentclass{article}
\usepackage{spconf,amsmath,graphicx,amsfonts,amssymb,amsthm,bm,bbm}
\usepackage{flushend}
\usepackage{enumitem}
\usepackage{cite}
\usepackage{color}
\usepackage{epstopdf}

\renewcommand{\vec}[1]{\mathbf{#1}} 

\title{FDD MASSIVE MIMO CHANNEL SPATIAL COVARIANCE CONVERSION USING PROJECTION METHODS }
\name{L. Miretti$^{\star}$, R. L. G. Cavalcante, and S. Sta\'nczak\thanks{$^{\star}$L. Miretti is now with EURECOM, France.}}%
\address{Fraunhofer Heinrich Hertz Institute}
\begin{document}
\ninept
\setlength{\belowdisplayskip}{1pt}
\setlength{\belowdisplayshortskip}{1pt}
\setlength{\abovedisplayskip}{1pt}
\setlength{\abovedisplayshortskip}{1pt}

\maketitle

\begin{abstract}
Knowledge of second-order statistics of channels (e.g. in the form of covariance matrices) is crucial for the acquisition of downlink channel state information (CSI) in massive MIMO systems operating in the frequency division duplexing (FDD) mode. Current MIMO systems usually obtain downlink covariance information via feedback of the estimated covariance matrix from the user equipment (UE), but in the massive MIMO regime this approach is infeasible because of the unacceptably high training overhead. This paper considers instead the problem of estimating the downlink channel covariance from uplink measurements. We propose two variants of an algorithm based on projection methods in an infinite-dimensional Hilbert space that exploit channel reciprocity properties in the angular domain. The proposed schemes are evaluated via Monte Carlo simulations, and they are shown to outperform current state-of-the art solutions in terms of accuracy and complexity, for typical array geometries and duplex gaps.
\end{abstract}
\begin{keywords}
FDD Massive MIMO, covariance matrix, angular reciprocity, projection methods, subspace estimation.
\end{keywords}
\section{Introduction}
\label{sec:intro}

In this work we address the problem of estimating the downlink (DL) spatial channel covariance matrix $\vec{R}^d$ in massive MIMO systems that operate in the frequency division duplexing (FDD) mode. The availability of a reliable estimate of $\vec{R}^d$ is a key ingredient in solving the problem of DL channel state information (CSI) acquisition, which is indeed one of the main performance bottlenecks of large-scale array systems \cite{Caire1}. 

In particular, in the massive MIMO regime, conventional DL channel estimation techniques (e.g. the ones currently implemented in LTE) require a prohibitively large estimation overhead, expressed in terms of the length of pilot sequences. In extreme cases, the training time may even exceed the channel coherence time, making a reliable CSI acquisition not possible \cite{scalingMIMO}. Existing solutions are therefore typically based on downlink-uplink (UL) channel reciprocity of time division duplexing (TDD) systems. Owing to the small number of antennas at the terminals, UL channel estimates can be obtained within the channel coherence time. However, in FDD systems this reciprocity is in general not available, and the envisioned approaches to the CSI acquisition problem typically rely on the existence of a lower dimensional representation of the channel vector in the large-scale array regime. The approaches can be divided into two main categories: methods based on compressed sensing (CS) and methods based on second-order statistics. An example of the first category is a CS based technique with dictionary learning proposed in \cite{Rao}. Although promising, CS techniques do not take into account the space-time correlation properties of the channel, which are often modeled by the well-known WSS assumption (see Sect. \ref{sec:model}). In contrast, the approaches based on second order statistics exploit these properties, and they have been shown to reduce effectively the effort for DL CSI acquisition \cite{CorrMIMO,Caire2,Cottat}.

Since direct estimation of $\vec{R}^d$ by using DL training sequences is a very challenging problem, this work proposes a novel technique to infer $\vec{R}^d$ from the observed UL covariance $\vec{R}^u$, which is easier to estimate in practice. The proposed approach has also the benefit of eliminating continuous covariance feedback from the user equipment. Related state-of-the-art solutions in literature include: 
\begin{itemize}[nosep]
\item \cite{Splines} Resampling of $\vec{R}^u$ for a uniform linear array (ULA) at a different wavelength by using cubic splines.
\item \cite{Dict1} (and the follow-up study \cite{Dict2}) Interpolation of $\vec{R}^d$ from $\vec{R}^u$ and a dictionary of stored $(\vec{R}^d,\vec{R}^u)$ pairs measured at different UE locations.
\item \cite{FC} Definition of a frequency calibration matrix obtained via a truncated Fourier series representation of the so called \textit{angular power spectrum} (APS).
\end{itemize}
The main underlying assumption of the state-of-the-art techniques and of this work is the channel reciprocity in the angular domain, which is here modeled with the frequency invariance property of the APS (see Sect. \ref{sec:model}). Because of its crucial role establishing the connection between $\vec{R}^d$ and $\vec{R}^u$, the core part of this work is devoted to the development of an accurate technique for APS estimation given $\vec{R}^u$. Unlike related studies, we formalize the problem as a \textit{convex feasibility} problem, which enable us to apply very effective solutions based on projection methods in an infinite-dimensional Hilbert space. The resulting scheme is shown to outperform the existing solutions in several aspects (see Sect. \ref{sec:perf}). In fact, it achieves estimation accuracy and flexibility comparable to \cite{Dict1} (the most accurate and robust algorithm considered so far) but with complexity comparable to \cite{Splines} and \cite{FC} (simple dictionary-less approaches).

This study is structured as follows: In Sect. \ref{sec:model} we introduce the channel model. Two variants of the proposed algorithm are described in Sect. \ref{sec:algo}, while practical implementation aspects for uniform linear arrays (ULA) are detailed in Sect. \ref{sec:ULA}. In Sect. \ref{sec:perf} we evaluate the performance of the algorithms with numerical simulations, and we highlight the advantages of the proposed scheme over the competing approaches. 

\textit{Notation:} We use boldface to denote vectors and matrices.  $( \cdot)^T$ and $( \cdot )^H$ denote respectively the transpose and Hermitian transpose, and $\| \cdot \|_F$  the Frobenius norm. $L^2[I]$ denotes the set of all square Lebesgue integrable functions over the interval $I \subset \mathbb{R}$. Given a Hilbert space, we denote by $x^{(i)} \rightharpoonup x$ a sequence $(x^{(i)})_{i\in\mathbb{N}}$ weakly convergent to a point $x$. We use $\Re[\cdot]$ and $\Im[\cdot]$ to denote, respectively, the real and the imaginary parts. Throughout the paper, superscripts $(\cdot)^u$ and $(\cdot)^d$ indicate respectively UL and DL matrices, vectors, or functions when we need to emphasize the dependency on the carrier frequency.

\section{SYSTEM DESCRIPTION AND PROBLEM STATEMENT}
\label{sec:model}

For simplicity, we consider a flat-fading MIMO channel between a base station (BTS) with $N \gg 1$ antennas and a single-antenna user equipment (UE) in a 2D (azimuth-only) scenario. However, we point out that our results can be can be extended to more general settings.

By sampling the time-variant channel vector $\vec{h}(t)$ at intervals corresponding to the channel coherence period $T_c$, a classical channel model (e.g., see \cite{Caire2}) assumes $\vec{h}[k] := \vec{h}(kT_c)$, $k \in \mathbb{Z}$, to be a zero-mean WSS circularly-symmetric Gaussian process that is white in time, while being correlated in the spatial domain, such that
\begin{equation*}
\vec{h}[k] \sim \mathcal{C}\mathcal{N}(\vec{0},\vec{R}) \quad \text{ i.i.d. }.
\end{equation*}
In typical communication models (e.g., see the 3GPP technical report \cite{3GPP}), the spatial covariance matrix $\vec{R} \in \mathbb{C}^{N\times N}$ takes the form  
\begin{equation} \label{eq:R_expr}
\vec{R} = \int_{-\pi}^{\pi}\rho(\theta)\vec{a}(\theta)\vec{a}^H(\theta)d\theta,
\end{equation}
where the vector-valued function $\vec{a}:[-\pi, \pi] \to \mathbb{C}^N$, with the $n$th coordinate function denoted by $a_n$, describes the array response of the BTS for a given direction of arrival/departure (DoA/DoD) $\theta \in [-\pi, \pi]$; and $\rho:[-\pi, \pi] \longrightarrow \mathbb{R}^+$ denotes the so called \textit{angular power spectrum} (APS) determining the average received/transmitted power per unit angle. In the following, we assume $\rho$, $\Re[a_n]$, and $\Im[a_n]$ to belong to $L^2[-\pi,\pi]$. According  to  the WSS assumption, $\vec{R}$ is  invariant over  time. In  practice, it is a slowly-varying parameter, since real channels  can be  safely  assumed  to be WSS just over a certain window of time $T_{WSS}$, which in usual scenarios is several order of magnitude larger than the channel coherence time $T_c$ \cite{Cottat,Caire3}.

This paper addresses the problem of estimating the DL covariance matrix $\vec{R}^d$ from an observation of the UL covariance matrix $\vec{R}^u$. In FDD systems, $\vec{R}^d$ and $\vec{R}^u$ are different, but strongly linked through the APS which, unlike the array response, exhibits strong frequency correlation properties. Therefore, in what follows, we assume the APS to be frequency invariant, an assumption that holds for typical FDD duplex gaps (see \cite{MolishB,COST1} for further details). By this assumption and (\ref{eq:R_expr}), we can write $\vec{R}^u \in \mathbb{C}^{N\times N}$ and $\vec{R}^d \in \mathbb{C}^{N\times N}$ as
\begin{equation} \label{eq:R_U_expr}
\vec{R}^u = \int_{-\pi}^{\pi}\rho(\theta)\vec{a}^u(\theta)\vec{a}^u(\theta)^Hd\theta,
\end{equation}
\begin{equation}\label{eq:R_D_expr}
\vec{R}^d = \int_{-\pi}^{\pi}\rho(\theta)\vec{a}^d(\theta)\vec{a}^d(\theta)^Hd\theta.
\end{equation}
In the following, we assume that the array responses $\vec{a}^u$ and $\vec{a}^d$ are known; this knowledge is cell-independent and it holds for the entire lifetime of the antenna array. In Sect. \ref{sec:algo}, to simplify the description of the proposed solutions, we assume perfect knowledge of $\vec{R}^u$. However, later in Sect. \ref{ssec:ULA_Ru} and Sect. \ref{sec:perf} we drop this assumption.

To motivate our work, let us consider DL channel estimation in the massive MIMO regime. If no knowledge on second-order statistics is available, standard techniques based on orthogonal training sequences require pilots of length at least $N$. However, if $\vec{R}^d$ is known, this knowledge can be exploited to reduce the training overhead. More precisely, if $p := rank(\vec{R}^d) < N$ (see \cite{Cottat} for a detailed analysis of this condition) the Karhuen-Loeve transform yields
\begin{equation*}
\vec{h}[k] = \vec{U}_p\vec{\Delta}_p^{\frac{1}{2}}\vec{w}[k],
\end{equation*}
with $\vec{w}[k] \in \mathbb{C}^{p\times 1} \sim \mathcal{C}\mathcal{N}(\vec{0},\vec{I})$ i.i.d, and $\vec{U}_p\vec{\Delta}_p^{\frac{1}{2}}\in \mathbb{C}^{N\times p}$ corresponding to the $p$ non-zero eigenvalues and respective eigenvectors of the eigen-decomposition $\vec{R}^d = \vec{U}\vec{\Delta}\vec{U}^H$. So if $\vec{R}^d$ is known, then, as shown in \cite{CorrMIMO,Caire2}, it is possible to reduce the training sequence length from $N$ to $p$. Moreover, in \cite{Cottat}, this fact is exploited to reduce pilot contamination effects. 

\section{SPATIAL COVARIANCE CONVERSION VIA PROJECTION METHODS}
\label{sec:algo}

We now present two variants of a novel scheme for estimating $\vec{R^d}$ based on the knowledge of $\vec{R^u}$. By recalling the model and the assumptions introduced in Sect. \ref{sec:model}, the main idea can be summarized in two steps as follows:
\begin{enumerate}
\item Given $\vec{R^u}$, we obtain an estimate $\hat\rho$ of the APS from equation (\ref{eq:R_U_expr}), and by exploiting known properties of $\rho$.
\item We compute an estimate of $\vec{R^d}$ from (\ref{eq:R_D_expr}), by substituting $\rho$ with its estimate $\hat\rho$.
\end{enumerate}
In particular, in the first step, we formulate the APS estimation problem as a \textit{convex feasibility problem}. The aforementioned two proposed variants of the scheme differ from each other in the exploitation of the known properties of the APS, which leads to two algorithms with different accuracy-complexity trade-offs. 

\subsection{APS estimation via projection onto a linear variety (Algorithm 1)}
\label{ssec:CovTrans1}

We rewrite (\ref{eq:R_U_expr}) as a system of equations of the form
\begin{equation} \label{eq:DoA_system_of_eq}
    r_m^u = \int_{-\pi}^{\pi}\rho(\theta)g_m^u(\theta)d\theta \quad m = 1 \ldots M, \quad M = 2N^2,
\end{equation}
where $r_m^u \in \mathbb{R}$ is the $m$th element of $ \setlength{\arraycolsep}{1.2pt}
\vec{r}^u := \text{vec}(
    \begin{bmatrix}
        \Re\{\vec{R}^u\} & \Im\{\vec{R}^u\}\\
    \end{bmatrix})
$, and $g_m^u: [-\pi,\pi] \longrightarrow \mathbb{R}$ is the $m$th element of the corresponding vectorization of the matrix $\vec{a}^u(\theta)\vec{a}^u(\theta)^H$. In general, since covariance matrices are Hermitian, the number of different equations is at most $N(N-1)$. Notice that it is possible to modify the definition of the $\text{vec}(\cdot)$ operator such that all the duplicated equations of (\ref{eq:DoA_system_of_eq}) are removed, but for notation simplicity in this section this trivial operation is omitted.

Now let $\mathcal{H}$ be the Hilbert space of real functions in $L^2[-\pi,\pi]$ equipped with the inner product $\langle f,g \rangle =  \int_{-\pi}^{\pi}f(\theta)g(\theta)d\theta$. By Sect. \ref{sec:model}, $\rho$ and $g_m^u$ are members of $\mathcal{H}$, so that (\ref{eq:DoA_system_of_eq}) can can be written as
\begin{equation} \label{eq:DoA_system_of_eq_prod}
    r_m^u = \langle \rho, g_m^u \rangle \quad m = 1 \ldots M.
\end{equation}
The inverse problem of finding $\rho$ given $g_m^u$ and $r_m^u$, $m=1\ldots M$, is obviously ill-posed. However, by using the set-theoretic paradigm \cite{SetTheo1,SetTheo2,SetTheo3,SetTheo4}, we propose to estimate $\rho$ by solving
\begin{equation} \label{eq:conv_prob}
\text{find } \rho^* \in V:= \cap_{m=1}^M V_m \neq \emptyset,
\end{equation}
where $V_m := \{\rho \in \mathcal{H}: \langle \rho, g_m^u \rangle = r_m^u \}$ for $m=1\ldots M$.

Among all the possible solutions of (\ref{eq:conv_prob}) (all of which are equivalent given (\ref{eq:R_U_expr})), motivated by the low complexity algorithm implementation that we show in the following, we choose the minimum norm solution
\begin{equation*}
\hat{\rho} = \arg \min_{\rho^* \in V} \| \rho^* \|,
\end{equation*} 
which corresponds to the orthogonal projection $P_V(0)$ of the zero vector onto the linear variety $V$ \cite[Sect.~3.10]{ConvB}. This projection has the following well-known closed-form expression:
\begin{equation} \label{eq:sol_lin_var}
    \hat\rho(\theta) = \sum_{m=1}^M\alpha_mg_m^u(\theta),
\end{equation}
where $\bm{\alpha} := [\alpha_1 \ldots \alpha_M]$ is a solution to the linear system
\begin{equation} \label{eq:Gram_system}
    \vec{r}^u = \vec{G}^u\bm{\alpha},
\end{equation}
\begin{equation*}
 \vec{G}^u = 
    \begin{bmatrix}
        \langle g_1^u,g_1^u \rangle & \langle g_1^u,g_2^u\rangle &  \dots  & \langle g_1^u,g_M^u\rangle \\
        \langle g_2^u,g_1^u\rangle & \langle g_2^u,g_2^u\rangle &  \dots  & \langle g_2^u,g_M^u\rangle \\
        \vdots & \vdots &  \ddots & \vdots \\
        \langle g_M^u,g_1^u\rangle & \langle g_M^u,g_2^u\rangle &  \dots  & \langle g_M^u,g_M^u\rangle
    \end{bmatrix},
\end{equation*}
which is guaranteed to have at least one solution (notice that we do not assume linear independence of the $g_i^u$). Furthermore, from the projection theorem, all solutions lead to the unique projection $\hat\rho$. 

We obtain an estimate of $\vec{R^d}$ by replacing $\rho$ in the DL equivalent of (\ref{eq:DoA_system_of_eq_prod}) with its estimate $\hat\rho$ obtained in (\ref{eq:sol_lin_var}):
\begin{equation} \label{eq:linear_relation}
\hat{r}_m^d = \langle \hat\rho, g_m^d \rangle = \sum_{l=1}^M\alpha_l \langle g_l^u, g_m^d \rangle \quad  m = 1 \ldots M,
\end{equation}
which can be rewritten in matrix form as
\begin{equation*}
    \vec{\hat{r}}^d = \vec{Q}\bm{\alpha},
\end{equation*}
where $\vec{\hat{r}}^d$ is an estimate of the vector $ \setlength{\arraycolsep}{1.2pt}
\vec{r}^d := \text{vec}(
    \begin{bmatrix}
        \Re\{\vec{R}^d\} & \Im\{\vec{R}^d\}\\
    \end{bmatrix})
$, $\bm{\alpha}$ is a solution of the linear system ($\ref{eq:Gram_system}$) given the UL measurements $\vec{r}^u = \vec{G}^u\bm{\alpha}$ as mentioned above, and 
\begin{equation*}
 \vec{Q} = 
    \begin{bmatrix}
        \langle g_1^d,g_1^u \rangle & \langle g_1^d,g_2^u\rangle &  \dots  & \langle g_1^d,g_M^u\rangle \\
        \langle g_2^d,g_1^u\rangle & \langle g_2^d,g_2^u\rangle &  \dots  & \langle g_2^d,g_M^u\rangle \\
        \vdots & \vdots &  \ddots & \vdots \\
        \langle g_M^d,g_1^u\rangle & \langle g_M^d,g_2^u\rangle &  \dots  & \langle g_M^d,g_M^u\rangle
    \end{bmatrix}.
\end{equation*}
It is important to underline that both $\vec{G}^u$ and $\vec{Q}$ depend only on the array geometry, and they can thus be computed or measured only once for the entire system lifetime.

\subsection{Exploiting further properties of the APS (Algorithm 2)}
\label{ssec:CovTrans2}

In many applications, additional prior knowledge about the APS $\rho$ is often available (for example, support information). If this knowledge can be expressed in terms of closed convex sets, then it is possible to narrow the set of candidate solutions of (\ref{eq:conv_prob}) to obtain more accurate APS estimates. By separating the real and imaginary part of ($\ref{eq:R_expr}$), and by working in the space of real functions, the previous algorithm in Sect. \ref{ssec:CovTrans1} already implicitly takes into account the knowledge that $\rho$ is real valued. 
In the following, we propose an extension of the previous algorithm by taking into account that, being a power spectrum, $\rho$ is always non-negative. More precisely, we look at the problem 
\begin{equation} \label{eq:extended_conv_prob}
\text{find } \rho^* \in C:= V \cap Z,
\end{equation} 
where V is the linear variety defined in (\ref{eq:conv_prob}) and $Z = \{\rho \in \mathcal{H}: (\forall \theta \in [-\pi,\pi]) \quad \rho(\theta)\geq 0\}$ is the closed convex set of non-negative functions in $\mathcal{H}$. A solution to (\ref{eq:extended_conv_prob}) can be found by applying one of the many existing iterative projection methods for convex feasibility problems available in literature. These methods typically produce a sequence $(\rho^{(i)})_{i\in\mathbb{N}} \subset \mathcal{H}$ such that $\rho^{(i)} \rightharpoonup \rho^* \in C$. In particular, we use the following fast iterative method called \textit{extrapolated alternating projection method (EAPM)}, given by \cite{ProjA}
\begin{equation} \label{eq:EAPM}
\rho^{(i+1)} = \rho^{(i)} + \nu K_i \left[ P_V(P_Z(\rho^{(i)})) - \rho^{(i)} \right] \quad (\forall i \in \mathbb{N}),
\end{equation}
where $\nu \in (0,2)$ is a step size, and $K_i$ is the extrapolation parameter defined as
\begin{equation*}
K_i = \begin{cases} 
\dfrac{\|P_Z(\rho^{(i)}) - \rho^{(i)} \|^2 }{\|P_V(P_Z(\rho^{(i)})) - \rho^{(i)}\|^2}, & \mbox{if } \rho^{(i)} \not\in Z \\ 
1, & \mbox{if } \rho^{(i)} \in Z 
\end{cases}.
\end{equation*}
The initial condition $\rho^{(0)} \in V$ can be arbitrary, and we choose $\rho^{(0)} = P_V(0)$, defined in Sect. \ref{ssec:CovTrans1}. The projection $P_V: \mathcal{H} \to V \subset \mathcal{H}$ onto $V$ is given by \cite[Chapter~3]{ConvB}
\begin{equation*} 
P_V(x) = x - \sum_{m=1}^M\beta_mg_m^u + P_V(0),
\end{equation*}
with $\bm{\beta} := [\beta_1 \ldots \beta_M]$ being a solution to the linear system $\vec{b} = \vec{G}^u\bm{\beta}$ where the $m$th element of $\vec{b}$ is given by $b_m = \langle x,g_m^u \rangle$ and $g_m^u$, $G^u$ are defined in Sect. (\ref{ssec:CovTrans1}). The projection $P_Z: \mathcal{H} \to Z \subset \mathcal{H}$ is given by \cite[p.~284]{SetTheo2}
\begin{equation*}
P_Z(x) = \begin{cases} 
x(\theta), & \mbox{for } x(\theta)\geq 0\\ 
0, & \mbox{otherwise} 
\end{cases}.
\end{equation*}

Now, by proceeding along the same lines as in Sect. \ref{ssec:CovTrans1}, an estimate of $\vec{R^d}$ can be obtained by 
\begin{equation*}
\hat{r}_m^d = \langle \hat\rho, g_m^d \rangle  \quad m = 1 \ldots M.
\end{equation*}
which has the same form as (\ref{eq:linear_relation}) except that $\hat\rho$ results from (\ref{eq:EAPM}). 

\begin{figure*}[ht!]
\begin{minipage}[b]{0.33\linewidth}
  \centering
  \centerline{\includegraphics[width=1\linewidth,trim = {6mm 0 6mm 11mm}]{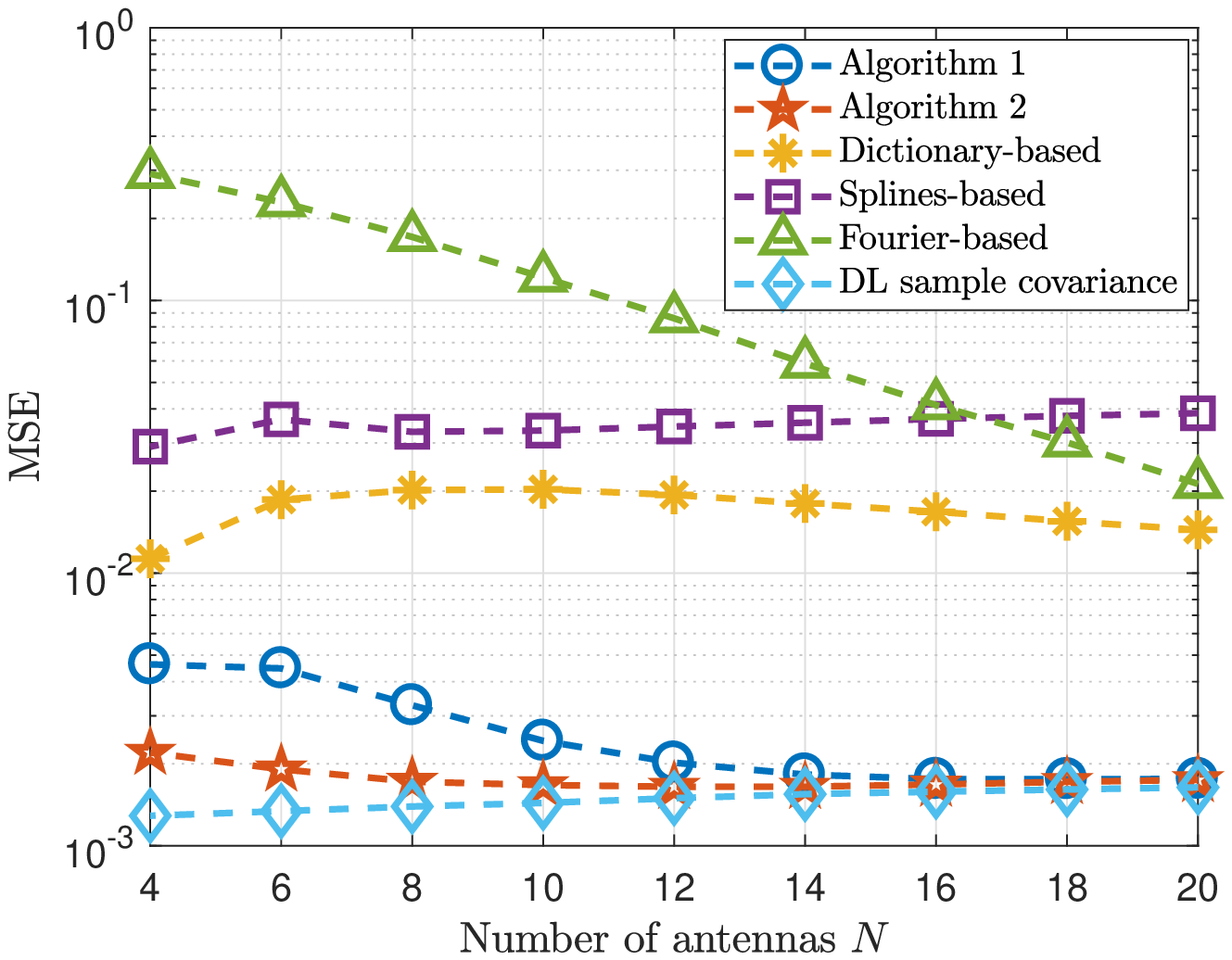}}
  \centerline{(a) Normalized Euclidean distance}\medskip
\end{minipage}
\begin{minipage}[b]{0.33\linewidth}
  \centering
  \centerline{\includegraphics[width=1\linewidth,trim = {6mm 0 6mm 11mm}]{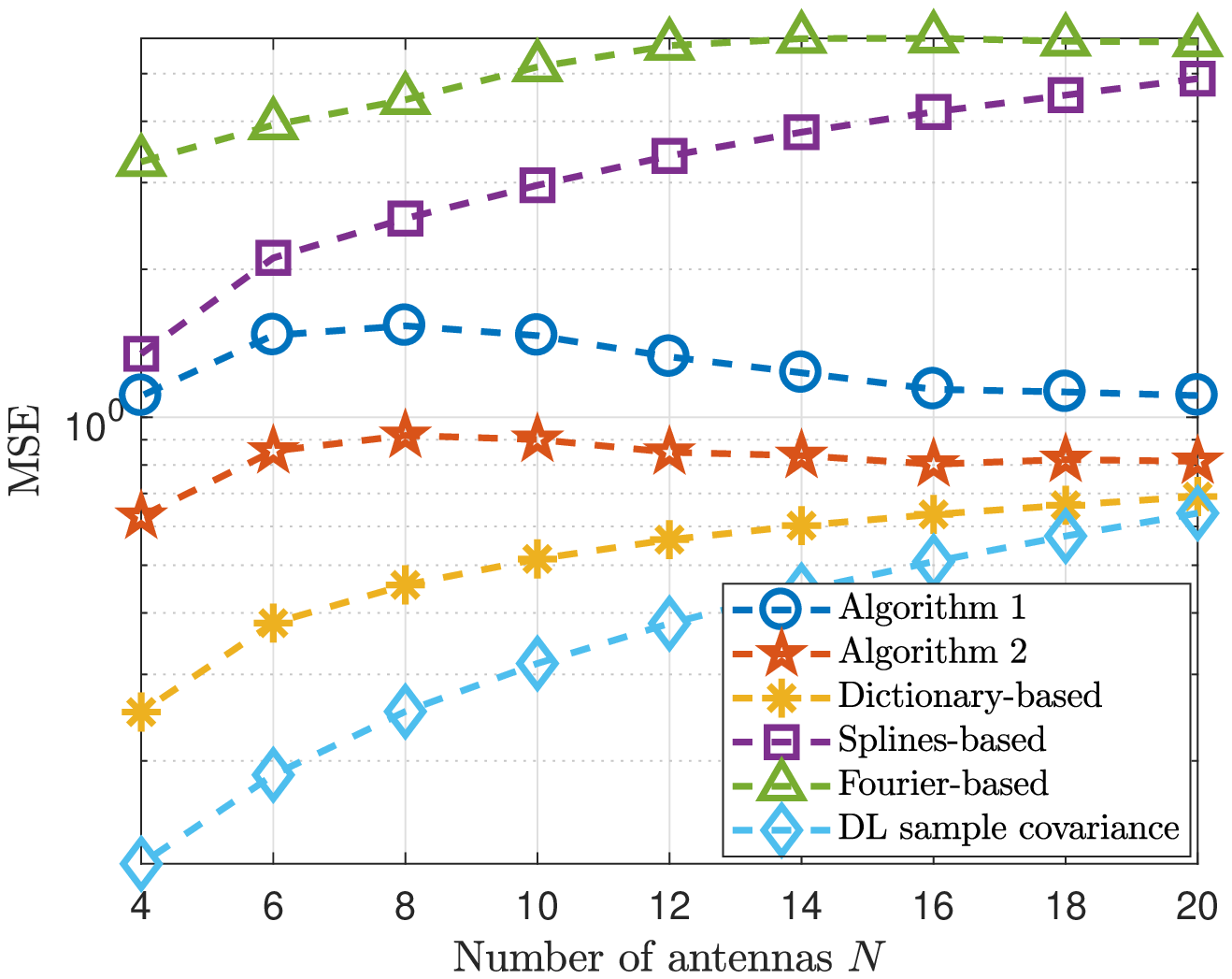}}
  \centerline{(b) Affine invariant distance}\medskip
\end{minipage}
\hfill
\begin{minipage}[b]{0.33\linewidth}
  \centering
  \centerline{\includegraphics[width=1\linewidth,trim = {6mm 0 6mm 11mm}]{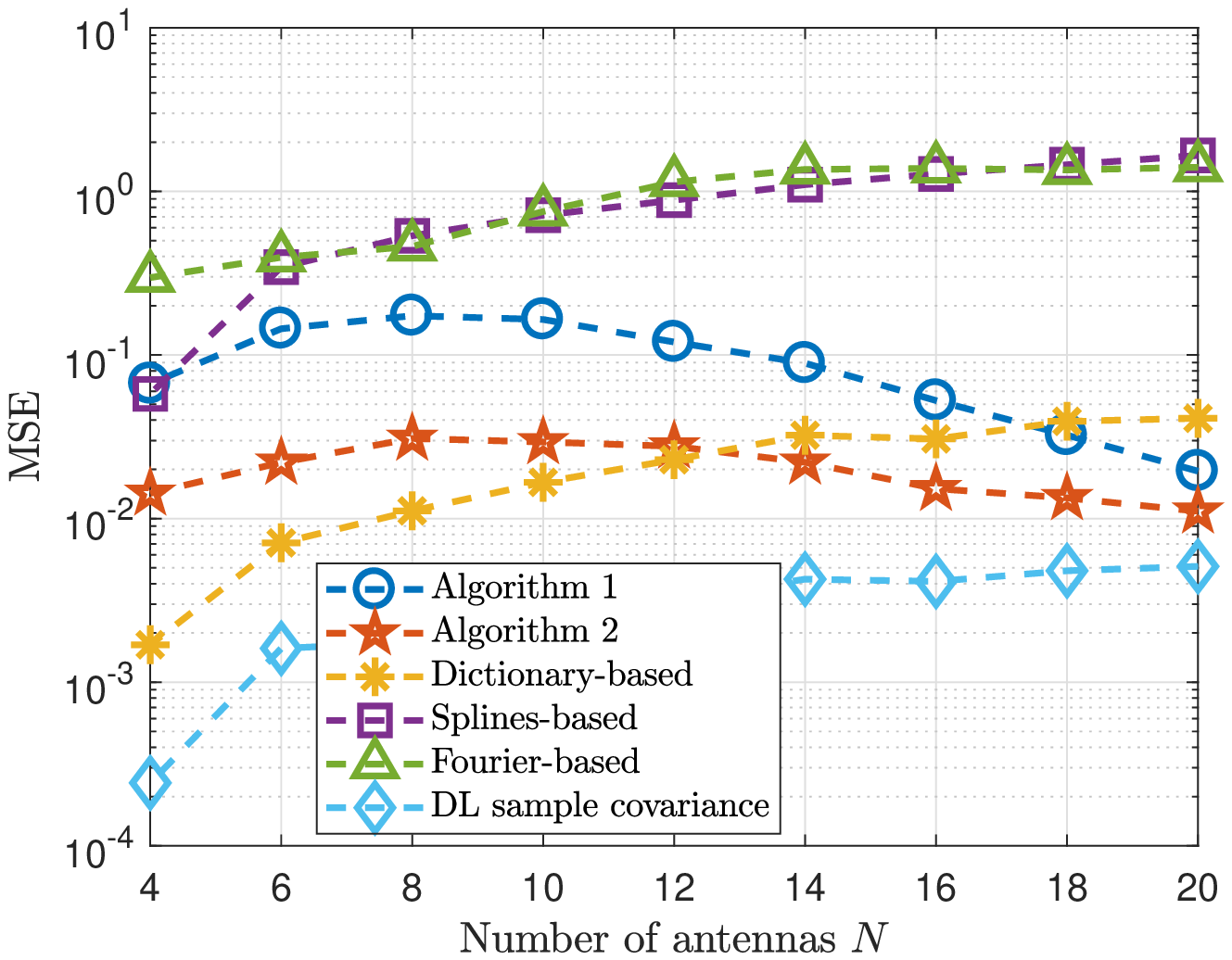}}
  \centerline{(c) Principal subspaces distance}\medskip
\end{minipage}
\caption{Simulation results: comparison of different DL covariance estimators vs number of BTS antennas N.}
\label{fig:comp}
\end{figure*}

\section{IMPLEMENTATION FOR UNIFORM LINEAR ARRAY}
\label{sec:ULA}

In this section we discuss an implementation of the proposed schemes to a uniform linear array (ULA) with $N$ antennas at the BTS. The array response of the ULA is given by
\begin{equation*}
\setlength{\arraycolsep}{2.5pt}
    \vec{a}(\theta) = \dfrac{1}{\sqrt{N}}
    \begin{bmatrix}
        1 & e^{j2\pi \frac{d}{\lambda}\sin\theta} &  \ldots  & e^{j2\pi \frac{d}{\lambda}(N-1)\sin\theta}\\
    \end{bmatrix}^T,
\end{equation*} 
where $d\in\mathbb{R}$ and $\lambda\in\mathbb{R}$ denote, respectively, the inter-antenna spacing and the carrier wavelength.

\subsection{Analytical expressions for $\vec{G}^u$ and $\vec{Q}$}
\label{ssec:ULA_matrices}

Since the ULAs are not able to distinguish among a DoA/DoD $\theta$ and its reciprocal $\theta + \pi$, we assume that the multipath components are confined to the interval $[-\pi/2, \pi/2]$, and we modify the definition of the scalar product for $\mathcal{H}$ accordingly, such that $\langle f,g \rangle =  \int_{-\pi/2}^{\pi/2}f(\theta)g(\theta)d\theta$. This assumption is supported by the fact that real systems often work with a similar or even narrower cell sectorization.

For ULA, the covariance matrix is positive semi-definite Hermitian Toeplitz, so it can be completely represented by its first column. By redefining $\text{vec}(\vec{A}) := \vec{a}_1$, where $\vec{a}_1$ indicates the first column of $\vec{A}$, we can prove that the matrices $\vec{G}^u$ and $\vec{Q}$ defined in Sect. \ref{ssec:CovTrans1} have the following analytical form expressed in terms of the Bessel function of the first kind, zero order $J_0: \mathbb{R} \to \mathbb{R}$:
\begin{equation*}
     \vec{G}^u = \dfrac{\pi}{2N^2}\begin{bmatrix}
        \vec{G}_\Re & \vec{0}       \\
        \vec{0}     & \vec{G}_\Im   \\
    \end{bmatrix} \quad
    \vec{Q} = \dfrac{\pi}{2N^2}\begin{bmatrix}
        \vec{Q}_\Re & \vec{0}       \\
        \vec{0}     & \vec{Q}_\Im   \\
    \end{bmatrix} ,
\end{equation*}
where the elements corresponding to the $(n,m)$-entries of $\vec{G}_\Re$, $\vec{G}_\Im$, $\vec{Q}_\Re$,$\vec{Q}_\Im \in \mathbb{R}^{N\times N}$ are given by
\begin{align*}
\begin{split}
    \vec{G}_{\Re,nm} = J_0(x_{nm}) + J_0(y_{nm}), &\ \vec{Q}_{\Re,nm} = J_0(p_{nm}) + J_0(q_{nm}),\\
    \vec{G}_{\Im,nm} = J_0(x_{nm}) - J_0(y_{nm}), &\  \vec{Q}_{\Im,nm} = J_0(p_{nm}) - J_0(q_{nm}),\\
\end{split}
\end{align*}
where
\begin{align*}
\begin{split}
x_{nm} =2\pi\frac{d}{\lambda^u}(n-m), &\quad p_{nm} =2\pi d\left(\frac{n-1}{\lambda^d} - \frac{m-1}{\lambda^u} \right), \\
y_{nm} =2\pi\frac{d}{\lambda^u}(n+m-2), &\quad q_{nm} =2\pi d\left(\frac{n-1}{\lambda^d} + \frac{m-1}{\lambda^u} \right). \\
\end{split}
\end{align*}
The proof is omitted because of the space limitation.

\subsection{Imperfect $\vec{R}^u$ knowledge}
\label{ssec:ULA_Ru}

In this section we analyze a scenario in which the BTS has access only to a UL sample covariance $\bar{\vec{C}}^u:= \frac{1}{K}\sum_{k=1}^K \hat{\vec{h}}^u[k](\hat{\vec{h}}^u[k])^H$ computed from a limited number $K$ of channel estimates defined as $\hat{\vec{h}}^u[k] = \vec{h}^u[k] + \vec{z}[k]$, $\vec{z}[k] \sim \mathcal{C}\mathcal{N}(\vec{0}$, $\sigma^2_z\vec{I})$ i.i.d. The noise power is obtained by setting a given per-antenna received $SNR := \frac{\mathbf{E}[|h_n|^2]}{\sigma_z^2}$. Let $\mathcal{H}_M$ be the Hilbert space of all $N\times N$ Hermitian matrices whose inner product is defined by $\langle A,B \rangle =  \text{trace}(B^HA)$, and let $\mathcal{C}$,$\mathcal{T}$ be the subsets of $\mathcal{H}_M$ composed respectively by positive semi-definite (PSD) and Toeplitz matrices.
The matrix $\bar{\vec{C}}^u$ is shown in \cite{Caire3} to be a sufficient statistic for estimating $\vec{C}^u := \mathbf{E}[\hat{\vec{h}}^u[k](\hat{\vec{h}}^u[k])^H] = \vec{R}^u + \sigma_z^2\vec{I} $, and in \cite{LMMSE} the matrix $\bar{\vec{R}}^u := \bar{\vec{C}}^u - \sigma_z^2\vec{I}$ is used to obtain the maximum-likelihood (ML)-PSD estimate of $\vec{R}^u$ by projecting it onto $\mathcal{C}$. The direct feeding of either $\bar{\vec{R}}^u$ or the ML-PSD estimate as input to the proposed algorithms may result in poor performance because the Toeplitz assumption imposed by the ULA is not satisfied. To overcome this problem, we propose to feed as input the projection of $\bar{\vec{R}}^u$ onto the set $\mathcal{T}_+:= \mathcal{C}\cap\mathcal{T}$, which imposes the desired Toeplitz structure. More precisely, we use $\hat{\vec{R}}^u = \arg \min_{\vec{X} \in \mathcal{T}_+} \| \vec{X} - \bar{\vec{R}}^u \|_F$ as the input. Since the projections on $\mathcal{C}$ and $\mathcal{T}$ are known \cite{PSDToepl} and easy to compute, it is possible to compute $\hat{\vec{R}}^u$ by applying standard methods such as the Dykstra's or Haugazeau's algorithm \cite[Chapter~29]{Dykstra}. In this work we use the approach described in \cite{PSDToepl}.

\section{SIMULATIONS AND FINAL REMARKS}
\label{sec:perf}
%

We simulate a typical model for the APS in cellular environments based on the well-known \textit{geometry-based stochastic channel model} (GSCM) \cite{MolishB}, where $\rho$ is assumed to be composed of a weighted superposition of probability density functions, supported by the intuition that the multipath components mainly originate from a set of $Q$ clusters of scatterers surrounding the BTS and the UE:
\begin{equation*}
\rho(\theta) = \sum_{q=1}^Qf_q(\theta)\alpha_q.
\end{equation*}
As an example, in the following we assume $Q$ uniformly drawn from $\{1,2,3,4,5\}$, Gaussian distributions $f_q \sim \mathcal{N}\left(\phi_q, \Delta_q^2 \right)$ with $\phi_q$ uniformly drawn from $[-\pi/3, \pi/3]$ and standard deviation (also called \textit{angular spread}) $\Delta_q$ uniformly drawn from $[3^{\circ},8^{\circ}]$, weights $\alpha_q$ uniformly drawn from $[0,1]$ and further normalized such that $\sum_{q=1}^Q\alpha_q = P_{RX}$, where $P_{RX}$ indicates the total received power. These statistical quantities are introduced to emulate the effect of different scattering patterns corresponding to random user locations.

A ULA is assumed for the BTS operating at UL/DL carrier wavelengths of $\lambda = 3\cdot10^8/f $ with $f = $ 1.8 Ghz and 1.9 Ghz respectively. The antenna spacing $d$ is set to half UL wavelength.
Channel realizations conditioned on $\vec{R}$ computed from (\ref{eq:R_expr}) are given by $\vec{h} = \vec{R}^{\frac{1}{2}}\vec{w}$, with $\vec{w} \sim \mathcal{C}\mathcal{N}(\vec{0},\vec{I})$. The BTS is assumed to have access only to a UL sample covariance matrix computed from $K =1000$ noisy channel estimates as described in Sect. \ref{ssec:ULA_Ru}.


The performance of the two algorithms defined in Sect. \ref{ssec:CovTrans1} and \ref{ssec:CovTrans2} are compared with the algorithms proposed in \cite{Splines}, \cite{Dict2}, and \cite{FC}, referred, respectively, to \textit{splines}-based, \textit{dictionary}-based, and \textit{Fourier}-based. The DL sample covariance, obtained with the same number of samples and SNR as for the UL, is used as a baseline. For fairness, all the sample covariances used in this comparison are corrected with the Toeplitzation procedure outlined in Sect. \ref{ssec:ULA_Ru}. The accuracy of an estimate $\hat{\vec{R}}$ of $\vec{R}$ is evaluated in terms of the mean square error $MSE :=
\mathbf{E}[e^2(\vec{R},\hat{\vec{R}})]$, where $e(\cdot,\cdot)$ is a given error metric. In particular, we consider:
\begin{itemize}[nosep]
\item The normalized Euclidean distance\\$e(\vec{R},\hat{\vec{R}}): = \| \vec{R}-\hat{\vec{R}}\|_F/\| \vec{R}\|_F$.
\item \cite{Dict1, CovMetrics} The affine invariant distance in the Riemannian space of PSD matrices $e(\vec{R},\hat{\vec{R}}): = \|\log(\vec{R}^{\frac{1}{2}}\hat{\vec{R}}^{-1}\vec{R}^{\frac{1}{2}})\|_F$.
\item \cite{CovMetrics} The Grassmanian distance between the principal subspaces $\vec{U}_p$,$\hat{\vec{U}}_p$ defined from  $\vec{R}$,$\hat{\vec{R}}$ by considering their eigenvectors corresponding to the minimum number $p$ of largest eigenvalues $\lambda_n$ satisfying $\sum_{n=1}^p \lambda_n / \sum_{n=1}^N \lambda_n \geq 95\%$. The metric is then $e(\vec{R},\hat{\vec{R}}): = \sqrt{\sum_{n=1}^p \gamma_n^2}$, where $\cos(\gamma_n)$ are the eigenvalues of $\vec{U}_p^H\hat{\vec{U}}_p$. This metric is particularly meaningful for the massive MIMO channel estimation problem, where a reliable signal subspace knowledge plays a crucial role.
\end{itemize}
The statistical mean is then obtained by Monte-Carlo simulations. For every Monte-Carlo run, a new APS and a SNR level $\in [10,30]$ (dB) are drawn.

Figure \ref{fig:comp} compares the algorithms for different numbers of BTS antennas $N$. Both the proposed algorithms approach the performance of the DL sample covariance estimator, as the number of constraints in the convex feasibility problem grows with $N$. The performance of both algorithms are comparable or better (depending on the metric and on the number of antennas) than the \textit{dictionary}-based method, which in principle can achieve extremely high accuracy given that the dictionary is sufficiently large (here we used 1000 entries). However, the proposed algorithms are dictionary-less, thus not requiring any overhead for dictionary acquisition. \textit{Algorithm 1} has the same low complexity of the \textit{Fourier}-based method, but it achieves much better accuracy. Compared to \textit{Algorithm 1}, \textit{Algorithm 2} shows better performance, especially in the low $N$ region, where the prior information about the positivity of the APS becomes important. However, the performance gains are achieved at the cost of higher complexity, which is due to the fact that the algorithm requires the numerical evaluation of integrals of the form $\int_{-\pi}^{\pi}x(\theta)d\theta$.

In summary, we have shown that the set-theoretic approach can be applied effectively to the problem of channel spatial covariance conversion. Compared to the competing approaches, the two variants of the proposed scheme are shown to achieve the high accuracy of the \textit{dictionary}-based method, but with the low complexity of the \textit{Fourier}-based and the \textit{splines}-based methods. 
\bibliographystyle{IEEEbib}
\bibliography{refs}
\end{document}